\documentstyle[aaspp,psfig]{article}

\def\Msun   {M$_{\odot}$}

\def\MIR    {10.8$\mu$m}

\def\CO     {CO(J$=$1$-$0)}

\def\sixcm  {$\lambda$6cm}

\def\etal{~{\it et al.}}

\def\Msun{~$M_{\odot}$}

\def\deg{\hbox{${}^\circ$}}

\def\las{\mathrel{\hbox{\rlap{\hbox{\lower4pt
        \hbox{$\sim$}}}\hbox{$<$}}}}
\def\gas{\mathrel{\hbox{\rlap{\hbox{\lower4pt
            \hbox{$\sim$}}}\hbox{$>$}}}}

\def\arcsec{\hbox{${}^{\prime\prime}$}}

\def\farcs{\hbox{$.\!\!{}^{\prime\prime}$}}

\begin{document}
\title{STRICT LIMITS ON THE IONIZING LUMINOSITY IN
NGC 1068 FROM JET-AXIS MOLECULAR CLOUDS}

\author{J. Bland-Hawthorn and S.L. Lumsden}
\affil{Anglo-Australian Observatory}
\author{G.M. Voit}
\affil{Johns Hopkins University}
\author{G.N. Cecil}
\affil{University of North Carolina}
\author{J.C. Weisheit}
\affil{Rice University}


\begin{abstract}
  The radio jet axis of NGC 1068 is characterised by energetic
  activity from x-ray to radio wavelengths. Detailed kinematic and
  polarization studies have shown that this activity is confined to
  bipolar cones centered on the AGN which intersect the plane of the
  disk. Thus, molecular clouds at 1 kpc distance along this axis are
  an important probe of the nuclear ionizing luminosity and spectrum.
  Extended \MIR\ emission coincident with the clouds is reasonably
  understood by dust heated to high temperatures by the nuclear
  radiation field. This model predicts that the nuclear spectrum
  is quasar-like (power law $+$ blue excess) with a luminosity 
  2$-$5 times higher than inferred by Pier\etal\ (1994). Consequently, 
  there is little or no
  polyaromatic hydrocarbon (PAH) emission associated with the
  radio-axis molecular clouds. We review this model in the light of
  new observations. A multi-waveband collage is included to illustrate
  the possible orientations of the double cones to our line of sight
  and the galaxian plane.
\end{abstract}

\section
{INTRODUCTION}

The luminous Seyfert 2 galaxy NGC 1068\footnote{Unless otherwise
  stated, global quantities for NGC 1068 are taken from the Ringberg
  Standard (Bland-Hawthorn\etal\ 1997).} lies at the forefront of
attempts to unify the broader class of ``active galaxies'' within a
single physical framework (Antonucci \& Miller 1985).  One of the
major uncertainties is the unobscured ionizing luminosity and spectrum
of the active nucleus. Fortuitously, the axes of the anisotropic
radiation cone and radio jet are inclined at roughly 45\deg\ to the
plane of the disk (Cecil\etal\ 1990; Gallimore\etal\ 1994), such that
the disk interstellar medium to the NE and SW sees the nuclear
radiation field directly. Notably, Pogge (1988) and Bergeron\etal\ 
(1989) find that highly ionized species (e.g. O$^{\scriptscriptstyle
  ++}$, Ne$^{\scriptscriptstyle ++}$, Ne$^{\scriptscriptstyle 4+}$) are
confined to a fan-shaped region aligned with the radio axis.

The inner 3 kpc region (Fig.~\ref{SixPix}$a$,$b$) is dominated by a
stellar bar aligned NE$-$SW (Scoville\etal\ 1988; Thronson\etal\ 1989)
and by molecular spiral arms which begin at the ends of the bar
(Planesas\etal\ 1991). In Fig.~\ref{SixPix}$a$, mid-infrared plumes
(Telesco \& Decher 1980) are coincident with the molecular clouds at 1
kpc radius, and have the same angular extent as the cones traced out
by the [OIII] emission (Fig.~\ref{SixPix}$c$) and by soft x-rays
(Fig.~\ref{SixPix}$d$).  Here, we investigate the possibility that the
extended mid-infrared emission arises from nuclear EUV/x-ray heating
of dust grains on the surface of molecular clouds.

\begin{figure}
  \begin{center}
    \leavevmode
    \psfig{file=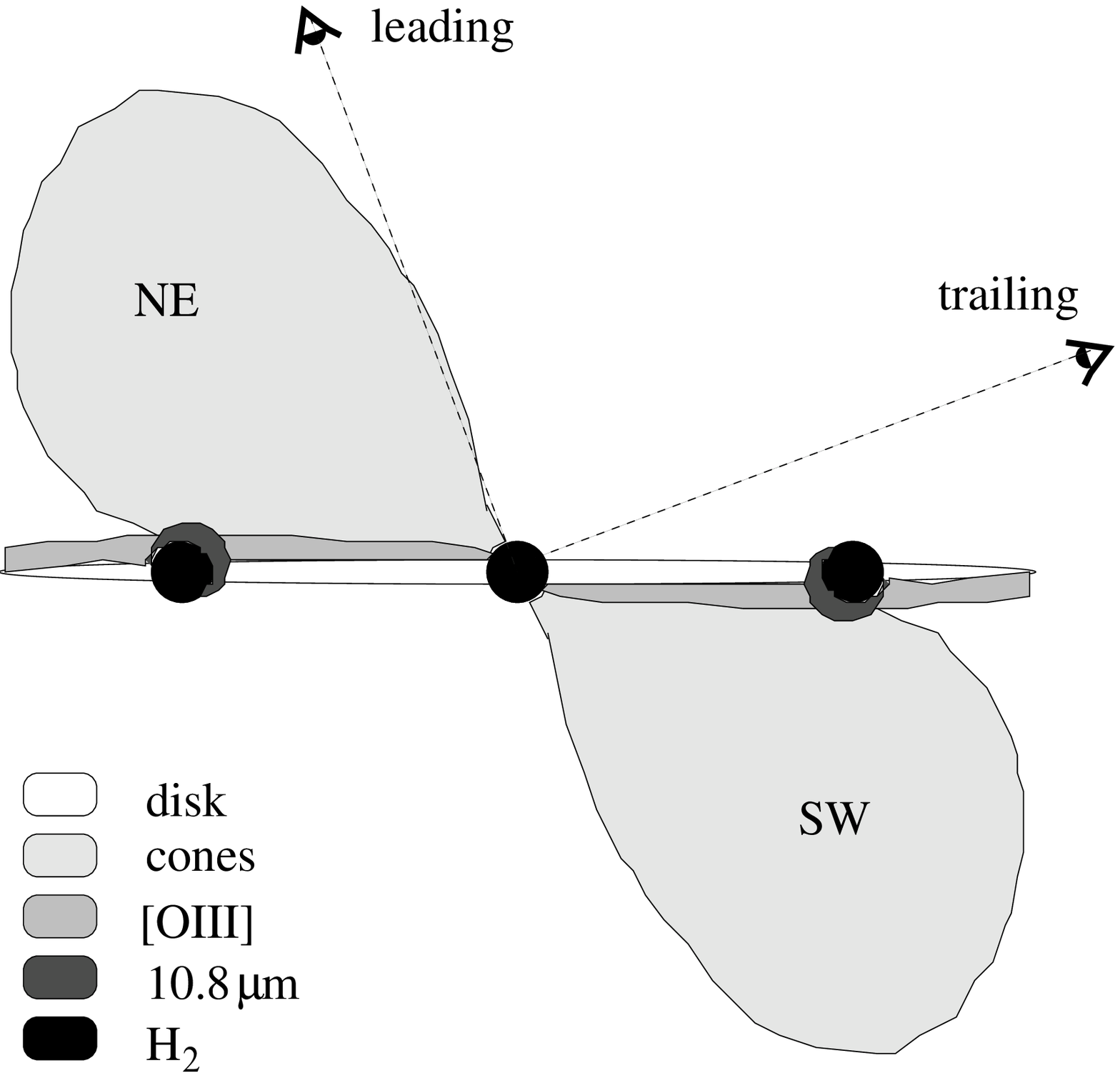,width=8.0truecm,height=6.0truecm}
  \end{center}
\caption{Schematic drawing of NGC 1068 in side elevation showing the
  orientation of the ionization/radio jet cones to the line of
  sight. If the spiral arms trail, the northern half of the disk falls
  behind the plane of the sky; if the spiral arms lead, the southern
  half falls behind. For either orientation, the NE cone lies in front
  of the disk, and the SW side behind the disk.}
\label{side-view}
\end{figure}

\begin{figure}
  \begin{center}
    \leavevmode
    \psfig{file=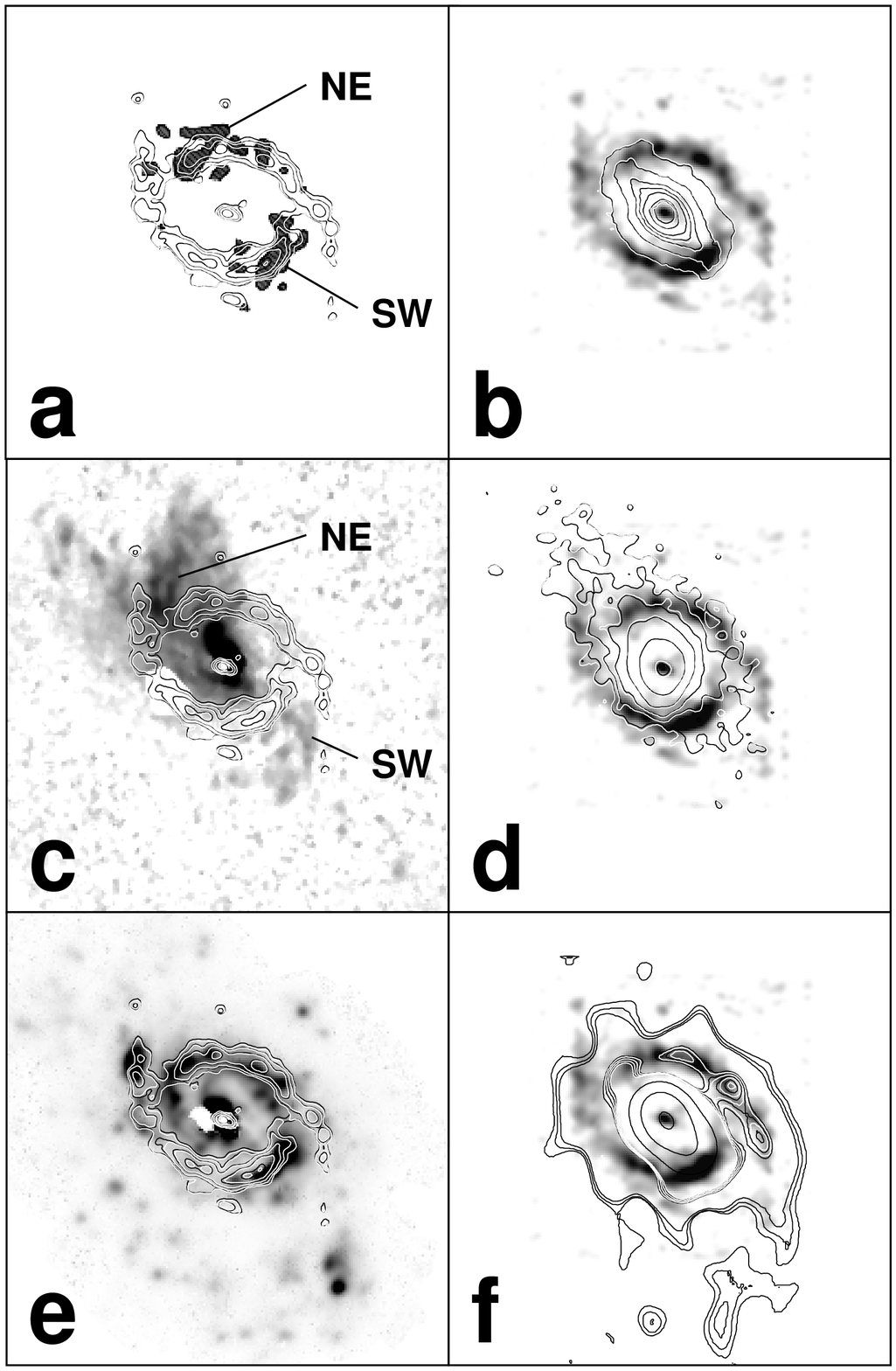,width=9.0truecm}
  \end{center}
\caption{A collage of multi-band observations over the inner
  145\arcsec\ of NGC 1068 superimposed on \CO\ line observations
  (Planesas\etal\ 1991) in contours ($a$, $c$, $e$) and in half tone
  ($b$, $d$, $f$). All maps are presented at 4\arcsec\ FWHM resolution
  except for the H$\alpha$ and [OIII] data shown at 1\arcsec\ FWHM
  resolution. In (a), the hatched regions are the \MIR\ emission
  observed by Telesco \& Decher (1988), i.e. the NE and SW 11$\mu$m
  plumes. In (b), the K band continuum from Thronson\etal\ (1989),
  shown with contours, reveals the central stellar bar. In (c), the
  half tone image is the [OIII]$\lambda$5007 line flux from
  observations with the Rutgers Fabry-Perot (the SE hole is an
  exorcised ghost). The NE and SW plumes do not coincide with the
  11$\mu$m plumes. In (d), the contours are the ROSAT (0.1$-$2.4 kev)
  HRI data (Wilson\etal\ 1994), where the flux is shared roughly
  between a point source and an extended (mostly NE) region. In (e),
  the half tone image reveals a dense distribution of HII regions in
  close association to the molecular gas. In (f), the contours are the
  VLA \sixcm\ data (Wynn-Williams\etal\ 1985): the NW radio
  `plateau' is resolved into a ridge coincident with the CO emission.
  }
\label{SixPix}
\end{figure}

\section
{THE ORIENTATION OF THE IONIZATION CONES TO THE GALAXY DISK}

\begin{figure}
  \psfig{file=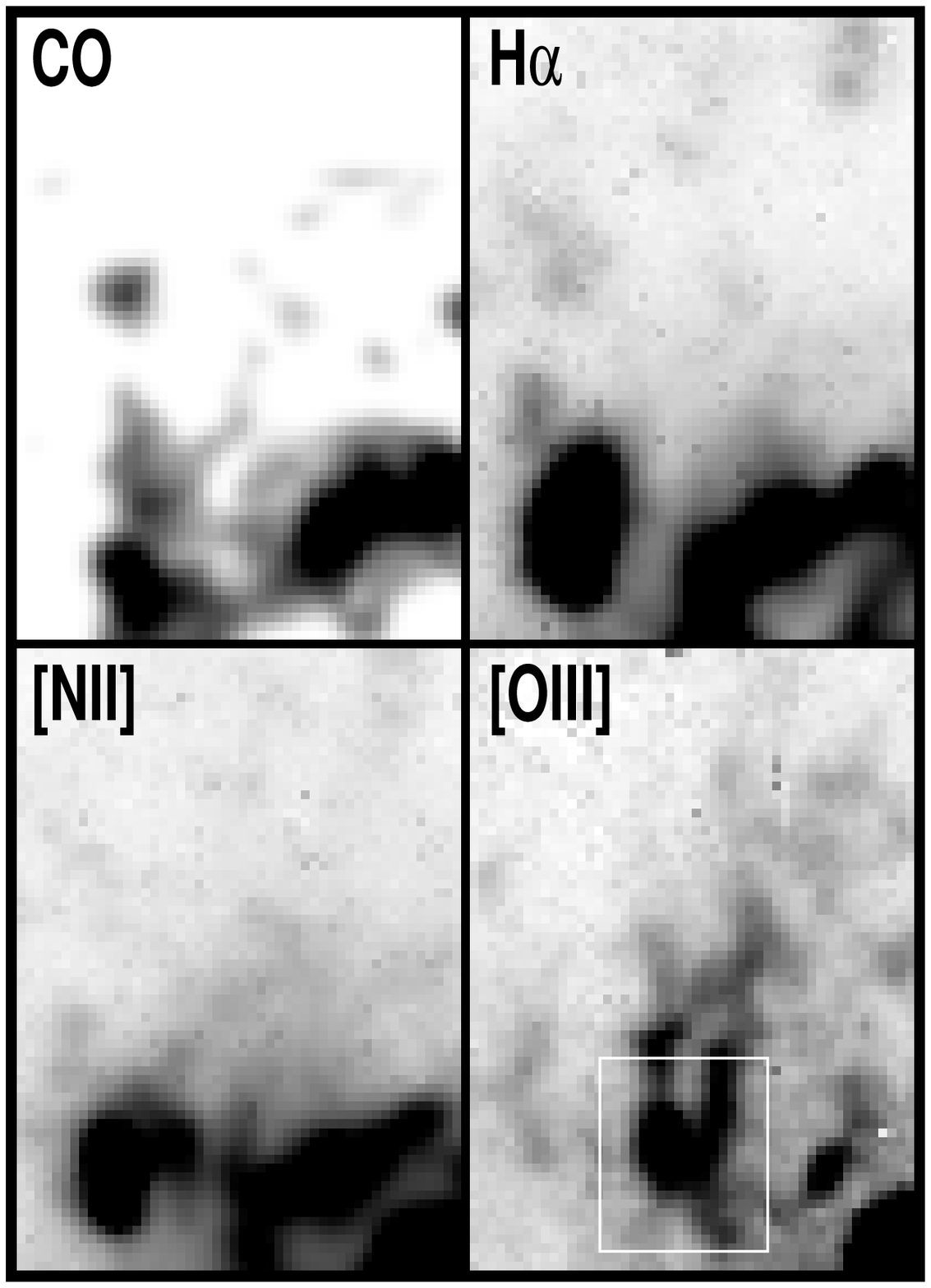,width=6.0truecm} \vskip -8.0truecm
  \hskip 6.0truecm \psfig{file=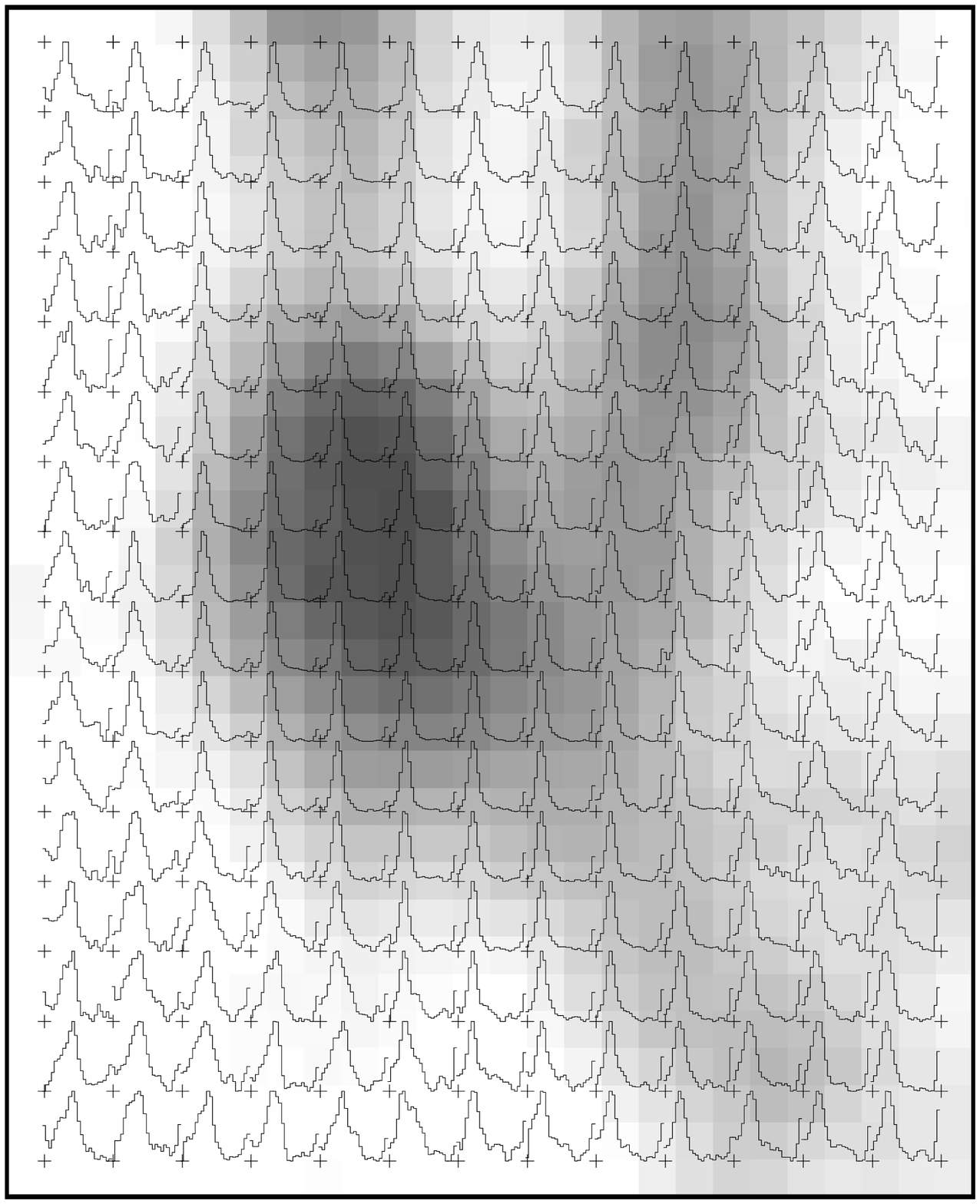,width=6.5truecm}
  \vskip 0.2cm
\caption{{\it Left:} (a) A comparison of the CO(J=1-0), [OIII],
  H$\alpha$ and [NII] flux in the NE [OIII] plume. {\it Right:}
  (b) The width of the [OIII] line profiles at the base of the plume
  correlates with the total line flux (shown in half tone) in the
  sense that the lines get broader as the flux decreases. Crosses
  indicate pixels binned 2$\times$2. The region shown is indicated
  by the white box in (a).}
\label{NE-plume}
\end{figure}

We emphasize that, like the double radio source, the ionizing photons
are escaping along bipolar cones. There is clear evidence that
the NE cone illuminates the nearside of the disk, whereas the SW cone
illuminates the farside, as illustrated in Fig.~\ref{side-view}.  In
Fig.~\ref{NE-plume}$b$, the line profiles throughout the bright NE
plume are narrow compared to line profiles in the diffuse emission
between plumes.  This is easily understood if the bright emission
arises from the surfaces of dense filaments, whereas the faint
emission is associated with the `diffuse ionized medium' (DIM) from a
vertically extended medium throughout the inner disk
(Bland-Hawthorn\etal\ 1991). In support of this picture, the [OIII]
plumes coincide with dense molecular gas (e.g.
Fig.~\ref{NE-plume}$a$). Moreover, the NE plume (Fig.~\ref{SixPix}$c$)
is remarkably similar to the SW plume if one makes allowances for
extinction by the disk ($A_V \approx 2$); their locations are exactly
bisymmetric with respect to the AGN. Note also how the [OIII] emission
to the SW is completely blocked by the CO ring.

The same phenomenon explains the asymmetry in the x-rays although the
extinction is more severe (Fig.~\ref{SixPix}$d$). To determine the
opacity in the ROSAT bandpass, we write
\begin{eqnarray}
  \label{tau_rosat}
  \tau_{\rm\scriptscriptstyle ROSAT} &=& \sum_{\rm\scriptscriptstyle ions} N_i \sigma_i(\varepsilon)\\
                              &=& A_V (420/\varepsilon)^3 p
\end{eqnarray}
where $p=[1+...]$ indicates higher order terms (Brown \& Gould 1976),
$i$ denotes different ions, $N_i$ is the ion column density, and
$\sigma_i$ is the absorption cross-section as a function of energy
$\varepsilon$ in eV. For the range 0.1$-$0.5 keV, He absorption
dominates the cross-section, while oxygen dominates in the range
0.5$-$2.4 keV; the bracketed terms are $p\approx3$ and $p\approx12$
respectively.  Thus, $\tau_{\rm\scriptscriptstyle ROSAT}$ lies in the
range $1.5 A_V$ to $3 A_V$. We suspect that the soft x-rays also arise
from filament surfaces as is well established in the M82 outflow (e.g.
Shopbell \& Bland-Hawthorn 1997).

\section
{THE MID-INFRARED EMISSION}

Telesco \& Decher (1988) found that almost all of the off-nucleus
\MIR\ emission occurs in two diametric `plumes' at roughly 15\arcsec\ 
radius along a NE$-$SW direction. The mid-infrared emission is clearly
associated with the dense CO arms (Planesas\etal\ 1991), and shows
some association with two of the brightest HII region complexes in the
region scanned by the bolometer (Fig.~\ref{SixPix}$e$). But the
excitation of these complexes is anomalous and shows evidence for
ionization from the central source (Evans \& Dopita 1986).  Moreover,
we note that strong \MIR\ emission is not detected in the NE complex
(radius $\approx$ 25\arcsec) nor in the NW complex at 10\arcsec\ 
radius, both of which fall within the scanned region. The NW complex
is the brightest source of UV emission outside of the circumnuclear
region (Neff\etal\ 1994).  Once again, we suspect that the MIR
emission arises mostly from the skin of the clouds.  Here, the NE and
SW plumes are similar in flux density and morphology because the
extinction is negligible at mid-infrared wavelengths.

From a reanalysis of VLA 6cm data (Wynn-Williams\etal\ 1985), we find
little correlation between the \MIR\ emission and the underlying
$\lambda\lambda 2-20$cm continuum (Fig.~\ref{SixPix}$f$).  When
compared with HII regions, the ratio of \MIR\ to radio cm bands is
high ($20-50$).  Thronson, Campbell \& Harvey (1978) find a tight
relation between 11$\mu$m and cm-wave flux densities for high surface
brightness galactic HII regions and molecular cloud complexes.  They
find S$_{\nu}$(11$\mu$m)/S$_{\nu}$(cm)$ = 10 \pm 3$ for a sample of
$\sim$40 objects.  In larger HII region complexes and starburst
galaxies, the synchrotron emission starts to overwhelm the free-free
emission at around $\lambda$2 cm (Condon \& Yin 1990), primarily
because the star forming regions are now sufficiently large that stars
are not able to migrate out of the region by the time they reach the
supernova stage.

\section
{DUST GRAIN HEATING}

Voit (1991; 1992) has considered how high energy photons heat and
evaporate dust grains in molecular clouds.  We now use the models to
examine the possible impact of the Seyfert nucleus on molecular gas
along the radio axis in NGC 1068.  Mathis, Rumpl \& Nordsieck (1977;
MRN) have presented a dust grain model that successfully explains the
`standard' extinction curve in terms of the absorption and scattering
properties of graphite and silicates. We adopt their power-law
distribution of grain sizes, $dn\ =\ (A_{\rm Si} + A_{\rm C})\ n_H\ 
a^{-3.5}\ da$, in the range 5\AA $\leq a \leq$ 0.5$\mu$m normalized to
the gas density, $n_H$.  The abundances for the silicates, $A_{\rm
  Si}$, and for graphite, $A_{\rm C}$, assumed to be 100\% carbon, are
given elsewhere (Voit 1991).

At a distance of 1 kpc from the AGN, the projected mass of gas
corresponds to $N_H \sim 1.2\times 10^{23}$ cm$^{-2}$. For a Galactic
dust to gas ratio, half of the total power in x-rays is absorbed by
the dust and then re-radiated in the infrared.  The remaining half
absorbed by gas-phase atoms excites an equivalent luminosity in UV
radiation which also heats the dust grains. Voit (1991) finds that
most of the grains are opaque to UV radiation, while being translucent
to x-rays. Since most of the surface area of the MRN composition lies
in the small grains, these are predominantly UV heated.  The x-rays
require a significant stopping column and therefore heat the large
grains preferentially. The temperature distribution of the grains
directly determines the form of the re-radiated spectrum. The smallest
grains, after absorbing individual photons, can superheat to
temperatures more than an order of magnitude higher than their
equilibrium temperatures (Puget \& Leger 1989) before cooling down
again. These `flickering grains' give rise to broad temperature
distributions.  X-ray photons can evaporate small grains ($a$ $\leq$
10\AA) on a very short timescale through repeated superheating to
temperatures in excess of 2000 K. EUV photons evaporate only the
smallest grains with sizes less than 6\AA\ (Jochim\etal\ 1994).

\section{ENERGY DISTRIBUTIONS}

Bland-Hawthorn \& Voit (1993) examine the influence of three different
ionizing continua: a simple power law and thermal bremsstrahlung
spectra at temperatures of 0.2 and 5 keV.  Here, we restrict our
attention to a power law and a more realistic `quasar spectrum'
incorporating a `big blue bump' (e.g. Sanders\etal\ 1989). For the
latter, we define
\begin{equation}
  \label{quasar}
  {\cal L} = k_1\ \varepsilon^{-2/3} \exp[-\varepsilon/\varepsilon_1] +
  k_2\ \varepsilon^{-\alpha} \exp[-\varepsilon/\varepsilon_2]\; {\cal H}(\varepsilon-\varepsilon_1)
\end{equation}
in units of photons per second per unit energy integrated over the
accretion disk, for which ${\cal H}$ is the Heaviside operator.  The
first term represents the `cool' accretion disk which produces the big
blue bump, for which the spectrum turns over at $\varepsilon_1 = 30$
eV (Fig.~\ref{Dust-models}$a$). The second term represents the
influence of the `hot' disk and corona, where the spectrum with slope
$\alpha = 1.9$ turns over at $\varepsilon_2 = 100$ keV. In practice,
$\varepsilon_2$ can be as high as 400 keV (Dermer \& Gehrels 1995).
The ratio of total energy radiated from the hot and cool regions,
$\rho$, is obtained from
\begin{equation}
  \label{k2}
  k_2 = {{L_{\rm tot}\ \varepsilon_2^{2-\alpha}}\over{(1+\rho)}}
  \left(\Gamma(2-\alpha) - 
  {{(\varepsilon_1/\varepsilon_2)^{2-\alpha}}\over{(2-\alpha)}} + 
  {{(\varepsilon_1/\varepsilon_2)^{3-\alpha}}\over{(3-\alpha)}} \right)^{-1}
\end{equation}
where $\Gamma$ is the Gamma function and $L_{\rm tot}$ is the total
luminosity from integrating equation~\ref{quasar} with respect to
energy. For our quasar nucleus, we adopt $\rho = 10$. 

The most crucial parameter in Voit's models is the x-ray flux incident
on the molecular cloud, $F_{\rm in} = \int f_{\nu} \ d\nu$.  We
consider flux levels of $10^8, 10^4$ and $10^2$ erg cm$^{-2}$ s$^{-1}$
equivalent to placing a molecular cloud at distances of 1 pc, 100 pc
and 1 kpc respectively from a 10$^{46}$ erg s$^{-1}$ ionizing source.
Further, we consider flux levels of $10, 1$ and 0.1 erg cm$^{-2}$
s$^{-1}$ for a cloud at 1 kpc from ionizing sources of $10^{45},
10^{44}$ and 10$^{43}$ erg s$^{-1}$ respectively.  The re-radiated
ionizing flux is given by
\begin{equation}
\label{flux}
F_{\rm out} = \kappa\left(\Delta\Omega/4\pi\right) \int_{\nu_0}^{\nu_1} f_{\nu} (1-e^{-\sigma_{\nu} N_H})  d\nu
\end{equation}
where $\kappa$ is the sky coverage fraction within the total solid
angle $\Delta\Omega/4\pi$ subtended by the molecular clouds, as seen
from the nucleus. If we assume a cylindrical geometry, the molecular
gas coincident with the \MIR\ emission lies within $\Delta\Omega/4\pi
= 0.1$; the enclosed mass of gas is roughly $1.5\times 10^{9}$\Msun\ 
with a mean density of 120 cm$^{-3}$. Here, we assume that giant
molecular clouds \#\#18--24 (Planesas\etal\ 1991) lie within a
cylinder of length 1.2 kpc and diameter 340 pc, such that the
projected column density seen from Earth is the same as that seen by
the AGN.

\section
{MODEL RESULTS}

In Fig.~\ref{Dust-models}$b$, we show the predicted spectra (solid
lines) from a simple power law where the limits of integration in
equation~\ref{flux} are taken to be $\nu_0$ $=$ 100 eV/h and $\nu_1$
$=$ 10 keV/h. The re-radiated spectrum is a strong function of the
x-ray flux incident on the molecular gas; the spectral shape is of
secondary importance.  The far-infrared emission provides a strong
constraint on an x-ray heating model: e.g., a flux of 40 erg cm$^{-2}$
s$^{-1}$ can explain the \MIR\ emission at the expense of predicting
too much far-infrared emission.  A flux level of 10 erg cm$^{-2}$
s$^{-1}$, equivalent to putting all of the bolometric luminosity
within our x-ray band, falls short by an order of magnitude in
explaining the mid-infrared emission.

There are two ways in which the re-radiated mid-infrared emission can
be enhanced at the expense of the far-infrared emission. First, if a
significant fraction of small grains ($\leq$ 10\AA) are able to
survive, their broad temperature distributions, while undergoing
thermal transients, can selectively enhance the near-infrared flux.
Secondly, the models discussed so far do not include a direct UV
component, nor do the models consider the UV and optical emission
expected to arise in the gas phase around the x-ray heated grains.  To
address the role of a more realistic quasar spectrum, we adopt the
form in equation~\ref{quasar} (Fig.~\ref{Dust-models}$a$) over the
interval 10 eV to 10 keV. The form of the ionizing continuum falls
within the observed range (Pier\etal\ 1994, Fig. 3), except beyond 2
keV where the observed x-ray spectrum turns up.  The predicted far
infrared spectra are illustrated with dashed lines in
Fig.~\ref{Dust-models}$b$.

At a flux level of 30 erg cm$^{-2}$ s$^{-1}$, such a model could
reasonably explain the \MIR\ detection at the same time as falling
within the limits imposed by the IRAS measurements.  The necessary
EUV ionizing luminosity is $3 L_{45}$ erg s$^{-1}$, where $L_{45}$
is the nuclear luminosity in units of 10$^{45}$ erg s$^{-1}$, a factor
of five larger than that inferred by Pier\etal\ (1994) for the
bolometric nuclear flux.  
  
The 25$\mu$m IRAS measurement poses a challenge for the `buried quasar'
model. However, an enhanced blue component is borne out by the
ISO observations of the narrow-line region for a wide range of ionic
species within the $2-40\mu$m window.  The range in ionization
potentials (30$-$300 eV) lead Lutz\etal\ (1997) to infer a strong
enhancement at EUV wavelengths. The additional ionizing flux we need
from the nucleus could
be reduced, and the 25$\mu$m IRAS constraint made less stringent, if
much of the UV is produced by hot young stars at the nucleus or
embedded within the CO arms.

There is always the possibility of autoionizing shocks driven by a
powerful $\sim 10^{56}$ erg wind associated with the radio jet.  In
Fig.~\ref{NE-plume}$a$, we note that the base of the NE [OIII] plume
radiates strongly in [NII] emission, and only weakly at H$\alpha$.
This situation routinely arises in the presence of fast shocks (Dopita
\& Sutherland 1995).  But in support of AGN photoionization of the
narrow-line region, Marconi\etal\ (1996) note that the infrared
[SiIX]/[SiVI] line ratio is very much higher than observed in
planetary nebulae and in gas surrounding hot young stars.

The factor $\kappa$ in equation~\ref{flux} allows for clumpiness
within the solid angle seen from the nucleus.  For a typical AGN
spectrum, $F_{\rm out}$ will depend much more sensitively on $\kappa$
than on $N_H$ at column densities of $10^{23}$ cm$^{-2}$. By way of
example, if we take $\kappa = 0.3$, increasing $N_H$ by a factor of 3
allows the cloud to absorb the x-ray power between about 3 keV and 5
keV, but this is a small fraction of the total power.  In the present
models, we do not consider self-absorption of the near- to
mid-infrared emission by the dusty medium. The details of this
correction depend on the cloud-source geometry and obviously depends
on whether we are viewing the irradiated or back surface of the
molecular cloud. For the measured column density, this is more
critical for the $K$ band where the extinction correction is expected
to be $\sim 10$ mag compared to $\sim 0.1$ mag at $\lambda$10$\mu$m.

\section{INFRARED EMISSION FROM SMALL GRAINS?}

\begin{figure}
\psfig{file=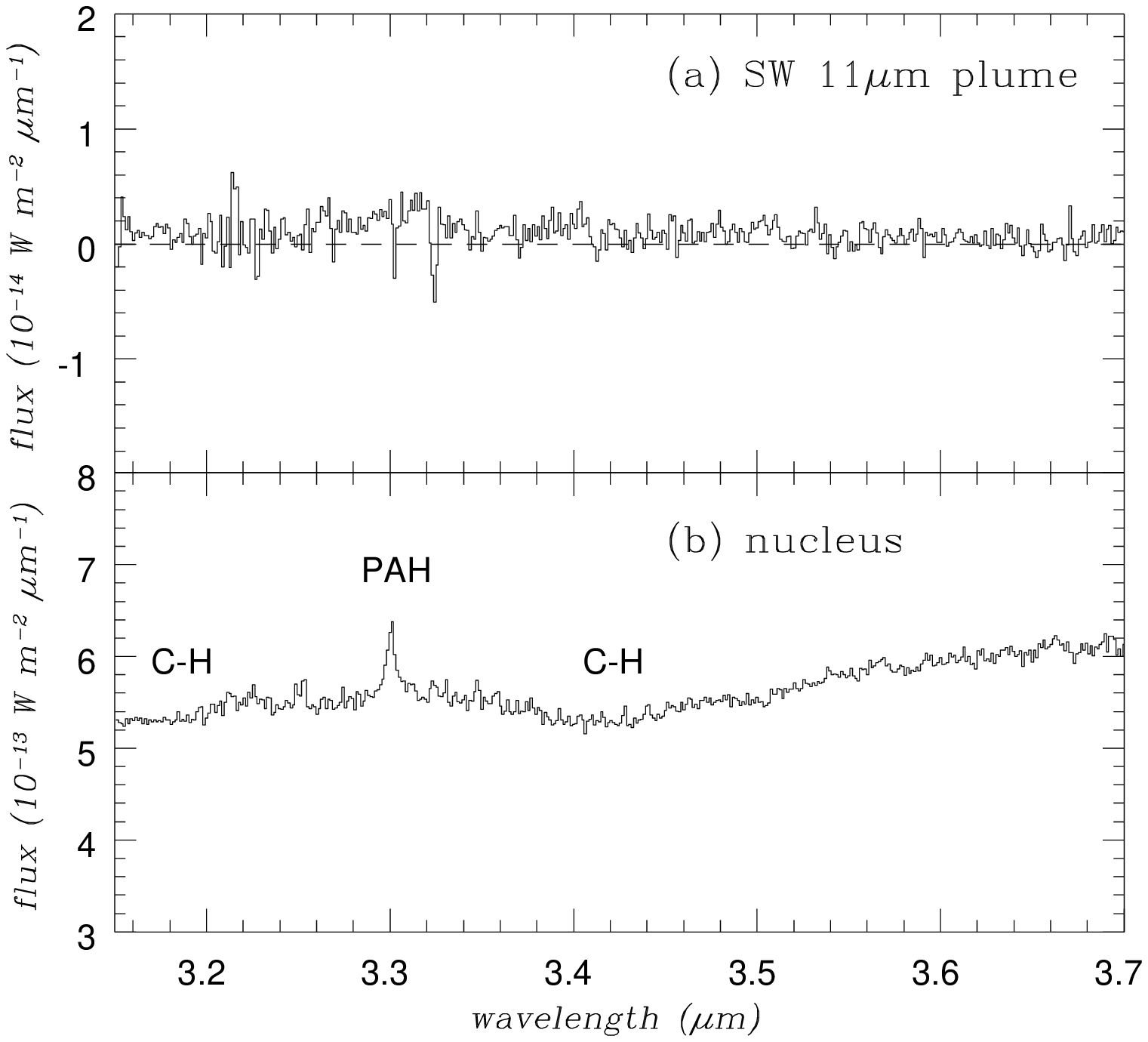,width=\textwidth,height=6.0truecm}
\caption{CGS4 spectra of the SW 11$\mu$m plume and the nucleus in NGC
  1068. (a) The identification of a nuclear PAH feature is uncertain
  (see text).  The C-H features indicate high absorption columns ($A_V
  \sim 20$) towards the nucleus.  (b) In addition to a weak continuum,
  there is a suggestion of weak PAH emission (broad base) in the SW
  plume.}
\label{3um}
\end{figure}

\begin{figure}
\centering
\hbox{
\ \psfig{figure=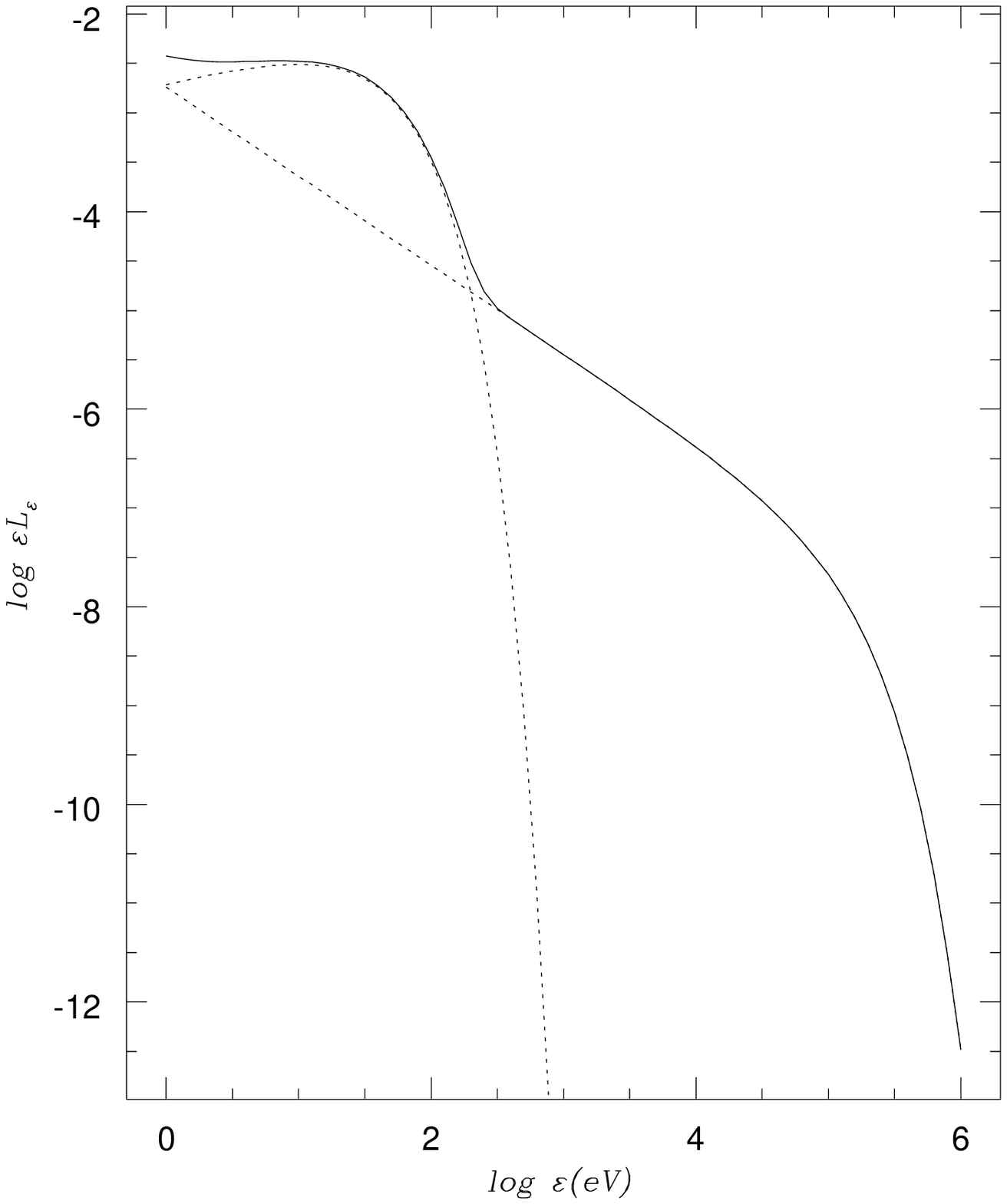,width=6.0truecm}
\ \psfig{figure=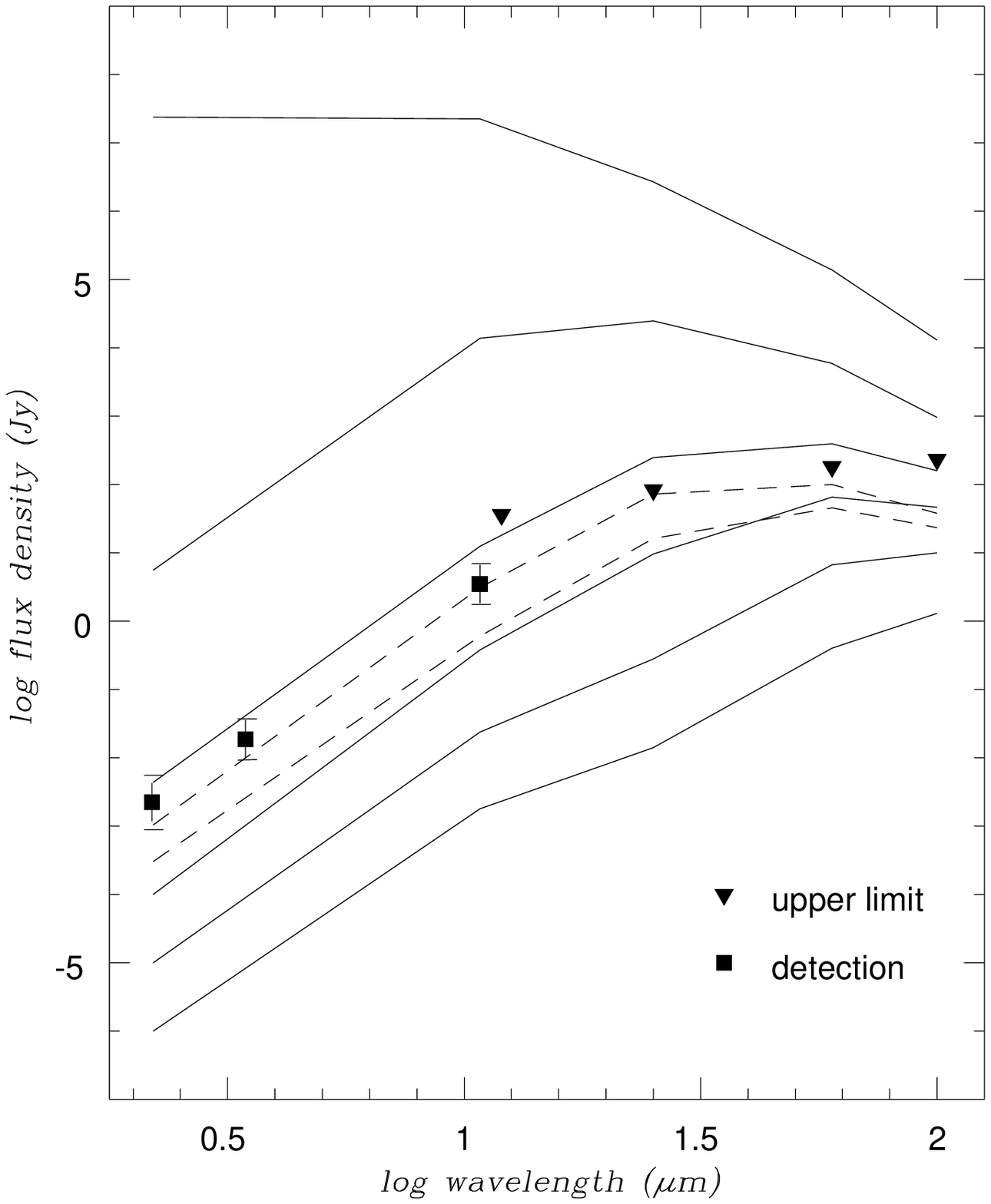,width=6.0truecm}
}
\caption{ {\it Left.} (a) Bolometric spectrum from the 
  accretion disk model where the total energy been normalized to
  unity. The `big blue bump' turns over at 30 eV; the hard energy
  component turns over at 100 keV.  {\it Right.} (b) The predicted
  near- to far-infrared spectra for the MIR features.  The solid lines
  are computed from the power-law models with flux levels of $10^8,
  10^4, 10^2, 10, 1, 0.1$ erg cm$^{-2}$ s$^{-1}$ (top to bottom).  The
  filled triangles indicate upper limits for the 11$\mu$m plumes from
  IRAS measurements. The filled squares are the detections discussed
  in $\S$7. The dashed lines correspond to flux levels of 10$^{1.5}$
  and 10 erg cm$^{-2}$ s$^{-1}$ and include a quasar-like `big blue
  bump' component.  }
\label{Dust-models}
\end{figure}

As noted by Wynn-Williams\etal\ (1985), an alternative explanation is
$\lambda\lambda 3.3$ $-$ $11.3\mu$m line emission from PAH molecules
(e.g. Molster\etal\ 1996).  However, for the x-ray fluxes considered
in our models, PAH grains should be destroyed (Voit 1992;
Bland-Hawthorn \& Voit 1993).  To test this idea, the SW 11$\mu$m
plume was observed in the $L$ band on 1995 July 16 with the UKIRT CGS4
infrared spectrometer at a resolving power of 700.  The InSb array
gave 1\farcs 5 pixels and the entrance slit projected to 3\arcsec
$\times$90\arcsec. The exposure time was 90 sec over an elapsed time
of 40 mins due to overheads in beam switching every 20 sec. The data,
presented in Fig.~\ref{3um}, were flux calibrated using BS 859 (A7 IV
star). Similar experiments were carried out with the CASPIR infrared
spectrometer at MSO 2.3m and at 10$\mu$m with CGS3 at UKIRT although
the data have lower signal to noise. The measured flux at 3$\mu$m is
shown in Fig.~\ref{Dust-models}$b$, multiplied by a factor of two
to account for both plumes. The $K$ measurement was obtained from
data supplied by H. Thronson after subtracting the stellar bar; the
10.8$\mu$ measurement is from Telesco \& Decher (1988).

In Fig.~\ref{3um}, while weak continuum is detected in the SW 11$\mu$m
plume, there is only a suggestion of a broad, weak 3.3$\mu$m PAH feature.
In contrast, the nuclear spectrum shows a strong narrow feature near
the expected wavelength for PAH emission bracketted by broad C-H
absorption features. However, it is not at all clear that this is the
correct identification. The Infrared Space Observatory (ISO) has shown
that PAH features have broad bases ($\sim 0.1\mu$m) which are not seen
in our nuclear spectrum (Molster\etal\ 1996).  Moreover, strong PAH
emission is always observed in star forming regions (e.g.
Acosta-Pulido\etal\ 1996) and rarely seen near AGNs (Roche\etal\ 1991;
cf.  Mazzarella\etal\ 1994).  The emerging picture is that both EUV
(Jochim\etal\ 1994; Allain\etal\ 1996$a$) and x-ray radiation (Voit
1992) destroy small (PAH) grains, although EUV photons only wipe out
grains with less than 50 carbon atoms ($a<6$\AA; Omont 1986).
Star-forming regions are strong PAH emitters because the EUV flux is
rapidly attenuated by the neutral gas, leaving sub-ionizing photons to
heat the small grains. {\it To remove all PAH-type grains requires
  x-rays}.

Voit (1992) has laid down strict requirements for PAH survival in the
presence of x-rays where PAHs are assumed to rebuild themselves as
efficiently as possible. The critical ionization parameter
$U^{\scriptscriptstyle\rm CR}$ (number density of photons to gas
atoms) is 10$^{-4}$ above which all PAH structures are destroyed; this
important limit is even lower if the PAHs are primarily ionized
(Allain\etal\ 1996$b$).  At a distance of 1 kpc,
$U^{\scriptscriptstyle\rm CR} \sim 0.1 L_{45}$ for a mean gas density
of $n_H = 10^2$ cm$^{-3}$. Thus, an x-ray luminosity of $L_X \sim
10^{42}$ erg s$^{-1}$ is the minimum requirement for grain destruction
but only within a column of 10$^{22}$ cm$^{-2}$ (Voit 1992; eq. 7 and
Fig. 3). For $L_X$ a factor of ten higher, most of the PAH emitters
are expected to be wiped out.

Towards the nucleus, the C-H features indicate $A_V \approx 22$
(Bridger\etal\ 1994) but this corresponds to an optical depth of unity
in the $L$ band ($N_H \sim 4\times 10^{22}$ cm$^{-2}$). If the nuclear
line at $\lambda 3.3\mu$m is indeed due to PAH grains, we suspect that
these regions are shielded from the central accretion disk by much
higher columns of obscuring material or by shadowing in a highly
inhomogeneous medium. If the PAH emission is produced in gas with $n_H
\sim 10^6$ cm$^{-3}$ at a distance of 30 pc from the nucleus, a
shielding column of 10$^{24}$ cm$^{-2}$ $-$ presumably from the torus
$-$ is sufficient to block most of the x-rays.

\section{THE ROLE OF THE NUCLEAR IONIZING LUMINOSITY}

The Ringberg workshop constitutes a major landmark in the study of
active galactic nuclei.  The VLBA imaging (Gallimore\etal\ 1997) and
the most recent water maser results (Greenhill \& Gwinn 1997) appear
to suggest that the accretion disk about the central black hole (radio
source S1) is warped on parsec scales. This was originally proposed by
Phinney (1989$a$,$b$) to explain the observation that the ionizing
cones and radio axes in Seyferts are randomly oriented with respect to
the rotation axes of the host galaxy. The newly discovered
radiation-driven warping instability would seem to provide the most
promising explanation for misalignments on parsec scales (Pringle
1996). An optically thick, planar disk (to both absorption and
emission), illuminated by a compact radiation source at its center, is
unstable to warping under the action of radiation pressure.

Maser kinematics suggest that the mass of the central black hole is
$1-2\times 10^7 \ M_\odot$, implying that the nuclear bolometric
luminosity of $0.6 L_{45}$ erg s$^{-1}$ (Pier\etal\ 1994) is roughly
half the Eddington limit ($1-2 L_{45}$ erg s$^{-1}$). If a warped
accretion disk removes the need for a torus (Begelman 1997), there are
presently no reliable constraints on the unobscured ionizing flux.
The 11$\mu$m plumes suggest that the intrinsic EUV luminosity could be
significantly higher ($3 L_{45}$ erg s$^{-1}$) than the value inferred
by Pier\etal\ (1994). Pier attempts to infer the scattered fraction of
nuclear radiation in order to arrive at the true nuclear luminosity,
which leads to uncertainties of a factor of two. The nuclear
luminosity inferred from the 11$\mu$m emission could be lowered by as
much as 50\% if one considers the possible role of stellar heating
($\S3$).  Thus, our luminosity and that of Pier may be marginally
consistent.  For either value, it seems that radiation pressure is
likely to be important, at least in the innermost accretion flow, and
the accretion rate could be at super-Eddington levels. The expected
anisotropy of radiation from a super-Eddington accretion flow (Sikora
1981) could conceivably account for the degree of intrinsic beaming
required by the PAH emission and mid-IR reprocessed radiation.

In earlier papers, we demonstrated that the inner 10 kpc disk of NGC
1068 displays an isotropic, diffuse ionized medium with an implied
cooling rate of 0.2 $L_{45}$ erg s$^{-1}$ (Bland-Hawthorn\etal\ 
1991$a$, $b$).  Sokolowski\etal\ (1991) contend that the same
electron-scattering medium which gives rise to both the observed x-ray
spectrum and the polarized broad-line spectrum can produce a dilute,
hard ionizing continuum that is sufficiently energetic to power the
low ionization emission. Halpern (1992) suggests that the extended
x-ray disk observed by ROSAT (Wilson\etal\ 1992) may not be sufficient
to balance the cooling rate of the DIM, and reiterates that the active
nucleus could be an important source of ionizing radiation. A nuclear
EUV luminosity as high as $3 L_{45}$ erg s$^{-1}$ would be sufficient
to power the diffuse component. However, the lower value of Pier\etal\ 
(1994) requires a different source, presumably the extended disk.
Thus, establishing the true nuclear ionizing luminosity has important
consequences on all scales within NGC 1068.

\section{CONCLUSIONS}

The lack of PAH features in the SW 11$\mu$m plume probably indicates
that the x-ray flux at the distance of the CO arms is at least 0.1 erg
cm$^{-2}$ s$^{-1}$ which requires a nuclear x-ray luminosity of more
than $10^{43}$ erg s$^{-1}$. Since the observed luminosity is an order
of magnitude smaller (Monier \& Halpern 1987), the scattered x-rays
are insufficient to wipe out PAH grains from molecular clouds outside
of the ionizing cones. But future observations with better sensitivity
could show weak PAH features within the ionization cones, and these
would require a detailed understanding of the radiation transfer
through the molecular clouds.

If the jet-axis molecular clouds are primarily heated by the nucleus,
the infrared spectrum of the 11$\mu$m plumes indicates an unobscured
EUV ionizing luminosity as high as $3\times 10^{45}$ erg s$^{-1}$,
although this value could be reduced by up to 50\% if stellar heating
is also important within the jet-axis molecular clouds.  For a
typical `blue bump' to x-ray energy ratio of 30 (Sanders\etal\ 1989),
the unobscured x-ray luminosity would be 10$^{44}$ erg s$^{-1}$.  The
observed scattered x-rays would therefore require a Thompson depth,
$\tau_e = 0.01$, in rough agreement with theoretical models (q.v. 
Pier\etal\ 1994).

\noindent{\bf ACKNOWLEDGMENT.}
The authors are indebted to E.A. Pier and D. Lutz for invaluable comments
on an earlier manuscript.

\end{document}